# Dynamic phase diagrams of the Blume-Capel model in an oscillating field by the path probability method

**Mehmet Ertaş[*] and Mustafa Keskin**

*Physics Department, Erciyes University, 38039 Kayseri–Turkey*

We calculate the dynamic phase transition (DPT) temperatures and present the dynamic phase diagrams in the Blume-Capel model under the presence of a time-dependent oscillating external magnetic field by using the path probability method. We study the time variation of the average order parameters to obtain the phases in the system and the paramagnetic (P), ferromagnetic (F) and the F + P mixed phases are found. We also investigate the thermal behavior of the dynamic order parameters to analyze the nature (continuous and discontinuous) of transitions and to obtain the DPT points. We present the dynamic phase diagrams in three planes, namely (T, h), (d, T) and ($k_2/k_1$, T), where T is the reduced temperature, h the reduced magnetic field amplitude, d the reduced crystal-field interaction and the $k_2$, $k_1$ rate constants. The phase diagrams exhibit dynamic tricritical and reentrant behaviors as well as a double critical end point and triple point, strongly depending on the values of the interaction parameters and the rate constants. We compare and discuss the dynamic phase diagrams with dynamic phase diagrams that are obtained within the Glauber-type stochastic dynamics based on the mean-field theory and the effective field theory.
.

## I. INTRODUCTION

The Blume-Capel (BC) model is a spin-1 Ising model with bilinear (J) and a single-ion potential or crystal-field interaction (D) that was originally proposed, by Blume [1] and Capel [2] independently, to study magnetic systems. The BC model is one of the well known prototype models used to examine various physical systems, such as multicomponent fluids, ternary alloys, microemulsions, ordering in semiconductor alloys, electronic conduction models, metallic alloys, polymeric systems, proteins and liquid crystals (see [3] and references therein). The model has also played an important guiding role in the critical phenomena such as tricritical and re-entrant behaviors, multicritical special points and multicritical phase diagrams. These works and determining the equilibrium properties of the model were conducted by using well-known methods in equilibrium statistical physics, such as the mean-field approximation (MFA), high- and low-temperature series expansions,

---

[*] Corresponding author.
Tel: + 90 (352) 2076666#33134; Fax: + 90 (352) 4374931
E-mail address: mehmetertas@erciyes.edu.tr (M. Ertaş)

Green's function diagrammatic approach, effective-field theory (EFT), Monte Carlo (MC) simulations, Renormalization Group (RG) calculations and the cluster variation method (CVM), the nonperturbative approach based on a thermodynamic self-consistent and the expanded Bethe-Peierls approximation (see [3] and references therein). The exact solution of the BC model was also presented on the Bethe lattice by means of the exact recursion relations [4]. We should also mention that although the model was introduced about 40 years ago, its equilibrium behaviors are still actively investigated by taking different effects such as quenched randomness [5], random crystal-field [6] or by using interesting lattices, such as Archimedean lattices [7]. It has also been used to examine different systems or phenomena such as $(Fe_{0.65}Ni_{0.35})_{1-x}Mn_x$ and $Fe_pAl_qMn_x$ alloys [8], Fe-Al disordered alloys [9], the first-order phase transition [10] and the wetting transition [11].

On the one hand, although a great amount is known about the equilibrium properties of the BC model, on the other hand less is known about the dynamical aspects of the model. An early attempt to study the dynamical aspects of the BC model was made by Fiig et al. [12]. They investigated the decay of metastable states by using dynamical MC simulations. Manzo and Olivieri [13] used the model to study metastability and nucleation within dynamical MC simulations. Ekiz et al. [14] investigated the dynamics of the BC model by using the path probability method (PPM) [15] which is the natural extension into the time dependent of the cluster-variation method (CVM) [16]. In particular, they studied the "flatness" property of metastable states, the "overshooting" phenomenon as well as the role of the unstable state. Buendía and Machado [17] presented two dynamic phase diagrams of the model in the presence of a time-dependent oscillating external magnetic field while studying the kinetics of a classical mixed spin-1/2 and spin-1 Ising system in detail within the Glauber-type stochastic dynamics based on the mean-field theory (MFT) [18]. Keskin et al. [19, 20] studied the dynamic phase transition (DPT) in the BC model under the presence of a time-dependent oscillating external magnetic field and constructed the dynamic phase diagrams by using Glauber-type stochastic dynamics based on the MFT. Gulpinar and Iyikanat [21] investigated the model with quenched diluted single-ion anisotropy in the neighborhood of equilibrium states by irreversible thermodynamics methods. Recently, Deviren and Albayrak [22] studied the DPT in the kinetic BC model on the Bethe lattice. The magnetic dynamic properties of the model were studied within Glauber-type stochastic dynamics based on the EFT [23, 24].

In this paper, we employ the PPM to study the dynamic magnetic properties of the BC model under the presence of a time-dependent oscillating external magnetic field. In particular, we study the time variation of the average order parameters to obtain the phases in the system and the paramagnetic (P), ferromagnetic (F) and the F + P mixed phases are found. We also investigate the thermal behavior of the dynamic order parameters to analyze the nature (continuous and discontinuous) of transitions and to obtain the DPT points. We present the dynamic phase diagrams in three planes, namely (T, h), (d, T) and ($k_2/k_1$, T), where T is the reduced temperature, h the reduced magnetic field amplitude, d the reduced crystal-field interaction and $k_2$, $k_1$ the rate constants. We compare and discuss the dynamic phase diagrams with dynamic phase diagrams obtained within the Glauber-type stochastic dynamics based on the mean-field approximation [17, 19, 20] and the effective field theory [23, 24]. It is worthwhile mentioning that the DPT in nonequilibrium systems in the presence of an oscillating external magnetic field has attracted much attention in the past few decades, theoretically (see [17, 19, 23, 25] and references therein), analytically [26] and experimentally [27].

The outline of the paper is as follows. In Sec. II, the model and methodology (PPM) for the BC model in the presence of a time-dependent oscillating external magnetic field are presented briefly. The numerical results and discussions are given in Sect. III. A summary and discussion of the results are given in Sect. IV.

## II. MODEL AND METHODOLOGY

The BC model was originally introduced independently by Blume [1] and Capel [2] in quite different contexts. The Hamiltonian for the BC model is given by

$$\mathcal{H} = -J\sum_{\langle ij \rangle} S_i S_j - D\sum_i S_i^2 - H(t)\sum_i S_i, \qquad (1)$$

where the $S_i$ takes the values ±1 or 0 at each site i of a lattice and the summation index < ij > indicates a summation over all pairs of nearest neighbor sites. J is the bilinear exchange interaction and D is the crystal-field interaction parameters. The last term, H(t), is a time-dependent external oscillating magnetic field that is given by H(t) = $H_0$ cos(wt), and $H_0$ and w = 2πν are the amplitude and the angular frequency of the oscillating field, respectively.

The BC model is a three-state and two-order parameter system. The average value of each of the spin states will be indicated by $X_1$, $X_2$ and $X_3$, which are also called the internal, state or point variables. $X_1$, $X_2$ and $X_3$ are the average fractions of spins with values +1, 0 and −1, respectively. These variables obey the following normalization relation: $\sum_{i=1}^{3} X_i = 1$. Two long-order parameters are introduced as follows: The first one is the average magnetization $<S>$, which is the excess of one orientation over the other orientation, also called the dipole moment, and the second is the quadrupole moment Q which is the average squared magnetization$<S^2>$, i.e., $q \equiv <Q>$. They can be expressed in terms of the internal variables as

$$m \equiv <S> = X_1 - X_3 \quad \text{and} \quad q \equiv <S^2> = X_1 + X_3. \tag{2}$$

$X_i$ (i = 1, 2, 3) can be expressed as linear combinations of the average order parameters by using the normalization relation with Eq. (2)

$$X_1 = \frac{1}{2}(m+q),\ X_2 = (1-q)\ \text{and}\ X_3 = \frac{1}{2}(q-m). \tag{3}$$

Now, we employ the PPM to obtain the equations of motion or dynamic equations. In this method the rate of change of the state variables is written as [15]

$$\frac{dX_i}{dt} = \sum_{i \neq j}\left(X_{ji} - X_{ij}\right), \tag{4}$$

where $X_{ij}$ is the path probability rate for the system to go from state i to j and satisfies the detailed balance condition

$$X_{ij} = X_{ji}. \tag{5}$$

The following two options were introduced by Kikuchi [15]:

$$\text{(A)}\ X_{ij} = k_{ij} Z^{-1} X_i \exp\left(-\frac{\beta}{2}\left(\frac{\partial E}{\partial X_i} - \frac{\partial E}{\partial X_j}\right)\right), \tag{6a}$$

$$\text{(B)}\ X_{ij} = k_{ij} Z^{-1} X_i \exp\left(-\frac{\beta}{2}\left(\frac{\partial E}{\partial X_j}\right)\right), \tag{6b}$$

both of which fulfill the necessary requirements expressed by Eq. (5) and $\beta = k_B T$. Z represents the partition function that is defined as

$$Z = \sum_{i=1}^{3} e_i,\ e_i = \exp\left(-\frac{\beta}{2}\left(\frac{\partial E}{\partial X_i}\right)\right). \tag{7}$$

E is the internal energy and we obtain the internal energy in terms of the internal variables by solving Eq. (1) using Eq. (2)

$$E = -J(X_1 - X_3)^2 + D(X_1 + X_2) - JH(X_1 - X_3).  \tag{8}$$

Moreover, $k_{ij}$ are the rate constants with $k_{ij} = k_{ji}$ and the model contains two rate constants. The first rate constant is $k_{12} = k_{23} = k_1$ which is the insertion or removal of particles associated with the translation of particles through the lattices. The second is $k_{13} = k_2$ that is associated with the reorientation of a molecule at a fixed site. We give the occurrence of the rate constants in Table 1. We assume that double processes, the simultaneous insertion, removal or rotation of two particles do not take place; hence only single jumps are allowed. Kikuchi [15] called assumptions (A) and (B) recipes I and II respectively. We also note that the general behavior of the solution of the dynamic equations obtained by recipes I and II is not drastically changed [28]. Thus, we use recipe II and obtain the following set of coupled dynamic equations by using Eqs. (3), (4), (6b), (7), and (8), and found

$$\Omega \frac{dm}{d\xi} = \frac{-\left[2k_1 \cosh\left(\frac{1}{T}(m+h\cos\xi)\right) + \frac{k_1}{k}\exp\left(-\frac{d}{T}\right)\right]m + 2\frac{k_1}{k}(k-1)\sinh\left(\frac{1}{T}(m+h\cos\xi)\right)q + 2\frac{k_1}{k}\sinh\left(\frac{1}{T}(m+h\cos\xi)\right)}{2\cosh\left(\frac{1}{T}(m+h\cos\xi)\right) + \exp\left(-\frac{d}{T}\right)}, \tag{9a}$$

$$\Omega \frac{dq}{d\xi} = \frac{-\frac{k_1}{k}\left[2\cosh\left(\frac{1}{T}(m+h\cos\xi)\right) + \exp\left(-\frac{d}{T}\right)\right]q + 2\frac{k_1}{k}\cosh\left(\frac{1}{T}(m+h\cos\xi)\right)}{2\cosh\left(\frac{1}{T}(m+h\cos\xi)\right) + \exp\left(-\frac{d}{T}\right)}, \tag{9b}$$

where $m = <S>$, $q = <Q>$, $\Omega = \frac{w}{k}$, $\xi = wt$, $d = \frac{D}{2J}$, $h = \frac{H_0}{2J}$ and $T = (\beta J)^{-1}$. T, h, d, and $\Omega$ are dimensionless. Thus, one of the advantages of the PPM over Glauber-type stochastic dynamics based on the MFT [17, 19, 20] and the EFT [23, 24] is that we find the set of coupled dynamic equations for m and q. On the other hand, Glauber-type stochastic dynamics based on the mean-field approximation and the effective field theory gave the uncoupled dynamic equations for m and q. The other advantage is that more than one rate constant can be introduced in the PPM. We should also mention that Eqs. (9a) and 9(b) reduces to Eqs. (11) and (10) for $k_1 = k_2 = 1/\tau$ in Ref. 19 and Ref. 20, respectively. On the other hand, the disadvantage of the PPM over Glauber-type stochastic dynamics based on the EFT is that all the correlations are neglected, whereas the latter method considers the correlations [23, 24]. Solution and discussion of Eqs. (9a) and (9b) will be presented in the next section.

## III. NUMERICAL RESULTS AND DISCUSSION

### A- Phases in the system

The stationary solutions of Eqs. (9a) and (9b) give the phases in the system that will be a periodic function of $\xi$ with period $2\pi$; that is, $m(\xi+2\pi)=m(\xi)$ and $q(\xi+2\pi)=q(\xi)$. Moreover, they can be one of two types according to whether they have or do not have the property

$$m(\xi+\pi)=-m(\xi) \text{ and } q(\xi+\pi)=-q(\xi). \tag{10}$$

A solution satisfying Eq. (10) is called a symmetric solution which corresponds to a paramagnetic (P) solution and it exists at high values of T and h. In this solution, $m(\xi)$ and $q(\xi)$ always oscillate around the zero value and are delayed with respect to the external magnetic field. The second type of solution gives a ferromagnetic (F) phase that does not satisfy Eq. (10) and it is called a nonsymmetric solution. In this case, $m(\xi)$ and $q(\xi)$ do not follow the external magnetic field anymore; $m(\xi)$ and $q(\xi)$ oscillate around $\pm 1$ and $+1$ values, respectively. We solved Eqs. (9a) and (9b) by applying the numerical method of the Adams-Moulton predictor corrector method for a given set of parameters and initial values and these are presented in Fig. 1. Fig. 1(a) displays only the symmetric solution; hence the P phase occurs in the system, but Fig. 1(b) illustrates only the nonsymmetric solution, therefore we have the F phase. Neither solution depends on the initial values. In Fig. 1(c), both the F and P phases exist in the system, thus the system contains the F + P mixed phase.

### B- Reduced temperature dependence of dynamic magnetization and dynamic quadrupole moment

The dynamic magnetization (M) and dynamic quadrupole moment (Q) are defined as

$$M = \frac{1}{2\pi}\int_0^{2\pi} m(\xi)d\xi \text{ and } Q = \frac{1}{2\pi}\int_0^{2\pi} q(\xi)d\xi. \tag{11}$$

We solve Eq. (11) by combining the numerical methods of the Adams-Moulton predictor corrector with the Romberg integration and investigate the thermal behavior of M and Q for several values of interaction parameters. Few interesting results are presented in Fig. 2; hence the calculation of the DPT points and the dynamic phase boundaries among the phases are illustrated. In the figure, thick and thin lines represent M and Q, respectively. $T_c$ and $T_t$ are the second-order phase transition and first-order phase transition temperatures for both M and Q, respectively. In Fig. 2(a), if the initial values of M and Q are taken as 1.0, M = Q = 1.0 at zero temperature and M decreases to zero continuously as the reduced temperature (T) increases, therefore a second-order phase transition occurs at $T_c$ = 0.434, but Q decreases until $T_c$, and at $T_c$, it makes a cusp and then increases to Q = 2/3, because q = < Q > is defined as < $S^2$ >. One can define m = < S > as 3 < $S^2$ > - 2 and this definition ensures that Q = 0 at infinite temperature [29]. On the other hand, if the initial values M and Q are taken as 0, M = Q = 0 at zero temperature; as the T values increase M always becomes zero, but Q increases to the values of 2/3, shown as dashed lines in the figure. Therefore, below $T_c$ the F + P phase exists and the dynamic phase transition is from the F + P to the P phase. This fact is seen clearly in Fig. 3(d) for h = 0.18. Fig. 2(b) illustrates that the system undergoes a first-order dynamic phase transition at $T_t$ = 0.78, because M and Q show a discontinuous jump and the transition is from the F to the P phase which can be clearly observed in Fig. 3(f). Figs. 2(c) and (d) were obtained for h = 0.2, d = 0.25, k = 1.0, and $k_1$ = 0.1, and they illustrate that the system undergoes two successive phase transitions. The first is a first-order transition from the F phase to the F + P mixed phase at $T_t$ = 0.66 (Fig. 2(c)) and the second is a second-order phase transition from the F + P phase to the P phase at $T_c$ = 0.71 (Fig. 2(d)). These facts are seen clearly in Fig. 3(b) for h = 0.2. Figs. 2(e)-(f) were plotted for h = 0.02, d = -0.5, k = 1.0, and $k_1$ = 0.1 and they display that the system undergoes three successive phase transitions. The first two transitions are first-order phase transitions from the F + P phase to the F phase at $T_{t1}$ = 0.09 and from the F phase to the F + P phase at $T_{t2}$ = 0.53, seen in Fig. 2(e). The third dynamic phase transition is a second-order one from the F + P phase to the P phase at $T_c$ = 0.58, shown in Fig. 2(f). These three phase transitions are clearly seen in Fig. 3(c) for h = 0.02.

### C- Dynamic phase diagrams

The calculated dynamic phase diagrams were given in the (T, h) plane, as seen in Fig. 3. We also investigated the phase diagrams in the (d, T) and (k, T) planes, and constructed three interesting phase diagrams in the (d, T) plane, as illustrated in Fig. 4 and two in the (k, T) plane, shown in as Fig. 5 due to the reason that most phase diagrams in these planes can be readily obtained from the phase diagrams in the (T, h) plane, especially for very high and low values of h. In Figs. 3-5, the solid and dashed lines represent the second- and first-order phase transition lines, respectively. We denoted the dynamic tricritical point by a filled circle, and B is the dynamic double critical end point.

In Fig. 3, we find the six main different topological types of the phase diagrams for the various interaction parameters and observe the following five interesting phenomena. (i) The system exhibits two or one dynamic tricritical behavior, as seen in Fig. 2(a) and Figs. 2(b)-(d), respectively. (ii) The system also displays a reentrant behavior, namely for low values of T the P phase exists and a T value increases the possibility of the system passing from the P phase to the F + P phase, and then back to the P phase again, as seen in Fig. 3(e). (iii) For high values of h and low values of T, the system undergoes a first-order phase transition, but for low values of h and high values of T the second-order phase transition occurs. (iv) In general, the P, F, and F + P phases exist in the system. (v) For high values of d < 0 only the F+P and P phases occur (Fig. 3(e)), but for high values of d > 0 the F and P phases exist (Fig. 3(f)). (vi) For high values of d > 0 and d < 0, the system always undergoes a first-order phase line, as seen in Figs. 3(e) and (f).

In the (d, T) plane we obtain four main dynamic phase diagrams, as three of which cannot be readily obtained from the phase diagrams in the (T, h) plane, as seen in Fig. 4. In Fig. 4(a), the system displays a tricritical behavior and also the dynamic double critical end point (B). For low values of T and d the system undergoes a first-order phase transition and transition from the F phase to the P phase. For high values of T and d, the P, F, and F + P phases exist and the dynamic phase boundary between the P and F+P phase is a second-order phase line, but between the F + P and F phase it is a first-order line. The phase diagram in Fig. 4 (b) is similar to Fig. 4 (a), except for the following two differences. (i) The first-order phase line between the F + P mixed phase and F phase occur for high values of T and d becomes a second-order phase line. (ii) For low values of T and d, the F + P phase occurs and the phase boundary between the F + P and P phases is a first-order line; hence B point disappears and one more dynamic tricritical point appears. The phase boundary between the F + P and P phases is a second-order line for high values of T and d, but a first-order line for low values of T and d. Fig. 4 (c) is similar to Fig. 4(b), except for the following differences. For high values of T and d, the boundary between the F + P and F phases is a first-order line. Since this boundary is also a first-order line for very low values of T and d, the system contains one more tricritical point; hence it has three dynamic tricritical points.

We also investigated the dynamic phase diagrams on the (k, T) plane and can easily observe from Fig. 3 that the system always undergoes a first-order transition for high values of h, d > 0 and the transition is from the F phase to the P phase, but for d < 0 the transition is from the F + P phase to the P phase. The dynamic phase diagrams in Fig. 5 cannot be easily observed from Fig. 3. In Fig. 5(a), the system undergoes a second-order phase transition for high values of T and low values of k and a first-order phase transition for low values of T and high values of k; hence it also exhibits dynamic tricrital behavior. Moreover, a triple point (TP) also occurs that the F, F + P and P phases simultaneously exist at this point. Fig. 5(b) is similar to Fig. 5 (a), but the following differences are observed. For very low values of T, one more F + P phase exists and the dynamic boundary between the F and F + P phase is a first-order line. The tricrical and triple point disappear; the boundary between the P and F + P phase is always a second-order line, but that between the F + P and F phase is a first-order line.

## IV. SUMMARY AND CONCLUSION

We calculate the DPT temperatures and present the dynamic phase diagrams in the Blume-Capel model under the presence of a time-dependent oscillating external magnetic field by using the PPM. We study the time variation of the average order parameters to obtain the phases and investigate the thermal behavior of the dynamic order parameters to analyze the nature of transitions and to obtain the DPT points. We present the dynamic phase diagrams (T, h), (d, T) and (k, T) planes. The dynamic phase diagrams contain the P, F, F + P phases and exhibit the dynamic tricritical and reentrant behaviors as well as a double critical end point and triple point depending on interaction parameters.

By comparison, since PPM contains two rate constants, which are important for the investigation of the dynamic behaviors of systems, it gives more interesting and more main different topological phase diagrams than those obtained from Glauber-type stochastic dynamics based on the MFT [17, 19, 20] and the EFT [23, 24]. Moreover, the system exhibits reentrant behavior in the (T, h) plane within the PPM, but does not exhibit it within the previous two methods. It is also worthwhile mentioning that for $k_1 = k_2 = 1/\tau$ Eqs. (9a) and 9(b) reduces to Eqs. (11) and (10) in Ref. 19 and Ref. 20, respectively; hence the PPM gives exactly the same result as Glauber-type stochastic dynamics based on the MFT [17, 19, 20]. On the other hand, the disadvantage of the PPM over Glauber-type stochastic dynamics based on the EFT is that all the correlations are neglected, whereas the later method considers the correlations [23, 24]. However, the phase diagrams obtained from using Glauber-type stochastic dynamics based on the EFT are more similar to the phase diagrams constructed by Glauber-type stochastic dynamics based on the MFT, except that some of the first-order phase lines disappear within Glauber-type stochastic dynamics based on the EFT. Therefore, we conclude that the rate constants influence the dynamic behavior of the system very much.

Finally, we hope that this detailed theoretical study may stimulate further research to investigate the dynamic properties of more complex systems by using the PPM. We also hope it will shed some light on experimental research on the dynamical aspects of the systems.

## ACKNOWLEDGMENTS

This work was supported by Erciyes University Research Funds, Grant No. FBA-2012-4121.

**LIST OF THE FIGURE CAPTIONS**

**FIG. 1.** Time variations in the average magnetization (m) and quadrupole moment (q).

  (a) Exhibiting a paramagnetic phase (P), h = 0.5, d = 0.25, T = 0.4, k = 1.0, and $k_1$ = 0.1.

  (b) Exhibiting a ferromagnetic phase (F), h = 0.4, d = -0.25, T = 0.5, k = 1.0, and $k_1$ = 0.1.

  (c) Exhibiting a coexistence region or mixed phase (F + P), h = 0.3, d = -1.0, T = 0.2, k = 1.0, and $k_1$ = 0.1

**FIG. 2.** The reduced temperature dependence of the dynamic magnetization M (the thick solid line) and the dynamic quadrupole moment Q (thin solid line). $T_c$ and $T_t$ are the second- and the first-order phase transition temperatures for both M and Q, respectively.

(a) Exhibiting a second-order phase transition from the F + P mixed phase to the P phase for h = 0.18, d = -0.85, k = 1.0, and $k_1$ = 0.1; $T_c$ is found to be 0.434. Solid lines for the initial values M = Q = 1.0 and the dashed lines for M = Q = 0.0.

(b) Exhibiting a first-order phase transition from the F phase to the P phase for h = 0.2, d = 1.0, k = 5.0, and $k_1$ = 0.1; $T_t$ is found to be 0.78.

(c) and (d) Exhibiting two successive phase transitions for h = 0.2, d = 0.25, k = 1.0, and $k_1$ = 0.1. The first one is a first-order from the F phase to the F + P mixed phase at $T_t$ = 0.66, seen in Fig. 2(c) and the second one is a second-order phase transition from the F + P phase to the P phase at $T_c$ = 0.71, shown in Fig. 2(d).

(e) and (f) Exhibiting three successive phase transitions and plotted for h = 0.02, d = -0.525, k =1.0, and $k_1$ = 0.1; the first two transitions are first-order phase transitions from the F + P phase to F at $T_{t1}$ = 0.09 and from the F phase to the F + P phase at $T_{t2}$ = 0.53, seen in Fig. 2(e). The third dynamic phase transition is a second-order from the F + P phase to the P phase at $T_c$ = 0.58, shown in Fig. 2(f)

**FIG. 3.** Dynamic phase diagrams of the BC model in the (T, h) plane. The paramagnetic (P), ferromagnetic (F) and mixed phase (F + P) are found. Dashed and solid lines represent the first- and second-order phase transitions, respectively. The dynamic tricritical points are indicated with filled circles and B denotes the dynamic double critical end point.

(a) d = -0.25, k = 0.01, $k_1$ = 0.1, (b) d = 0.25, k = 1.0, $k_1$ = 0.1, (c) d = -0.5, k = 1.0, $k_1$ = 0.1, (d) d = -0.85, k = 1.0, $k_1$ = 0.1, (e) d = -1.0, k = 1.0, $k_1$ = 0.1, (f) d = 1.0, k = 5.0, $k_1$ = 0.1.

**FIG. 4.** Same as FIG. 3, but in (d, T) plane.

(a) h = 0.5, k = 0.5, $k_1$ = 0.1, (b) h = 0.1, k = 0.1, $k_1$ = 0.1, (c) h = 0.1, k = 5.0, $k_1$ = 0.01.

**FIG. 5.** Same as Fig. 3, but in (k, T) plane.

(a) h = 0.2, d = 1.0, $k_1$ = 0.1, (b) h = 0.05, d = -0.5, $k_1$ = 0.01.

**TABLE 1.** The description of the rate constants.

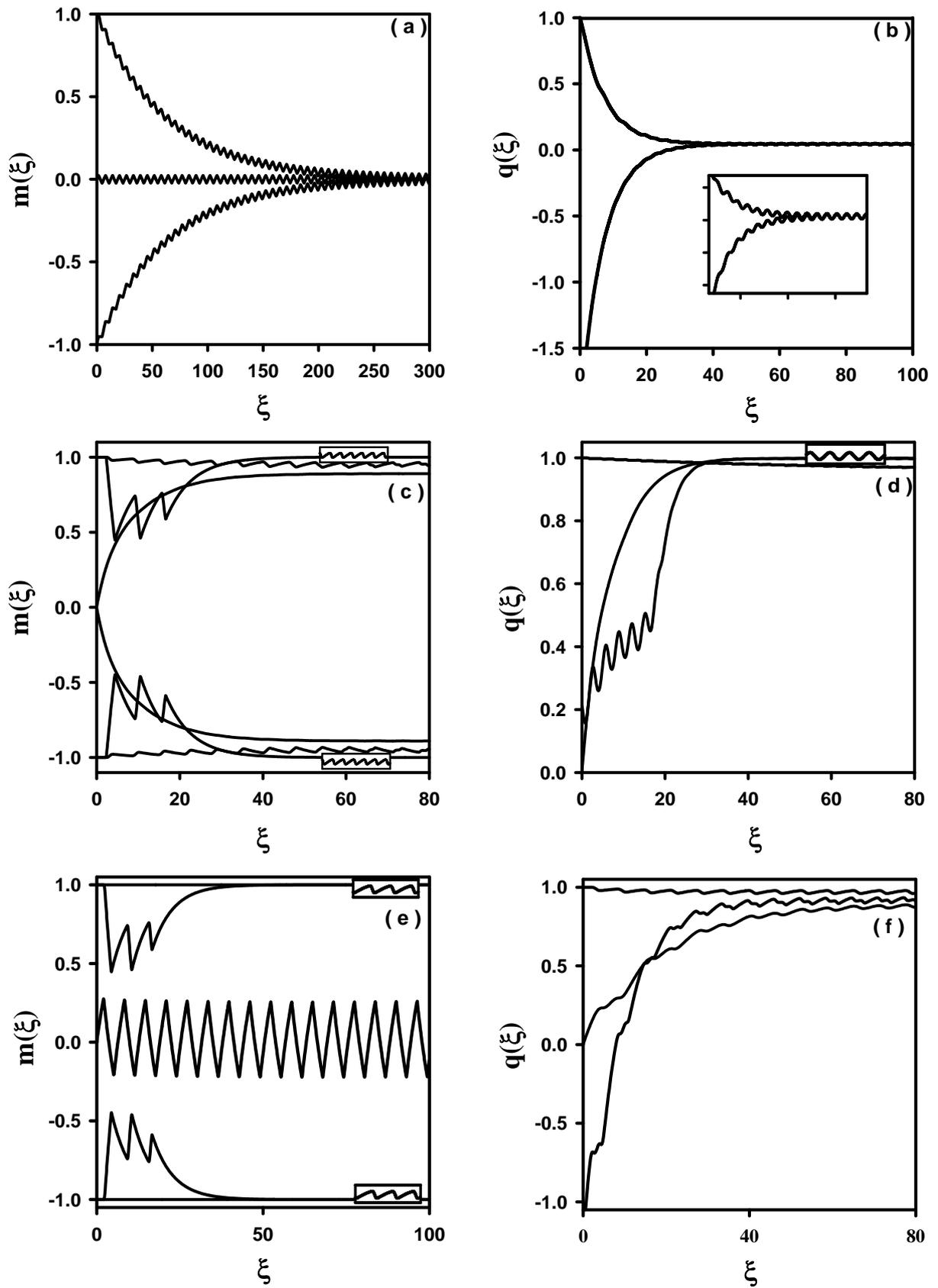

**FIG. 1**

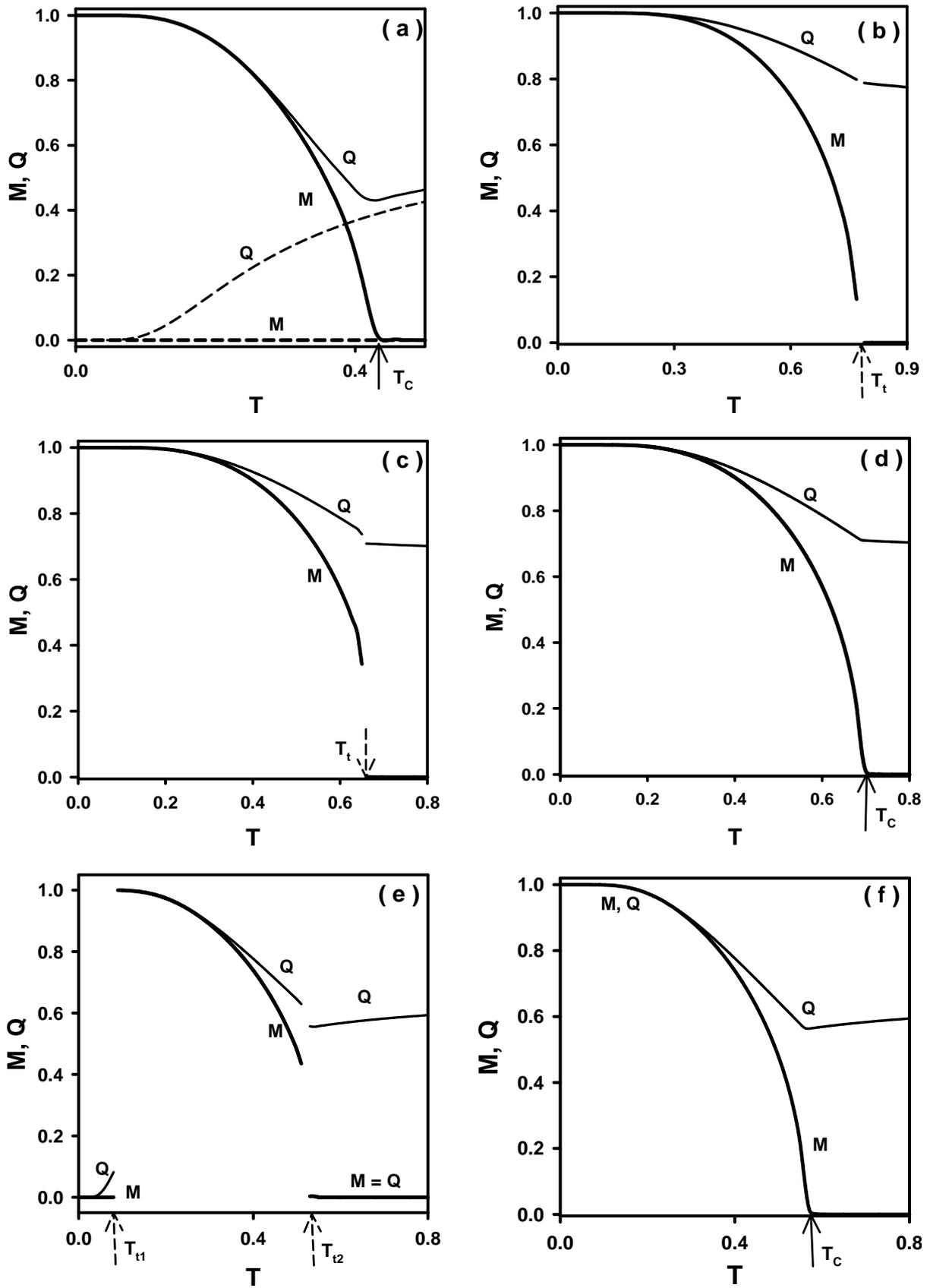

**FIG. 2**

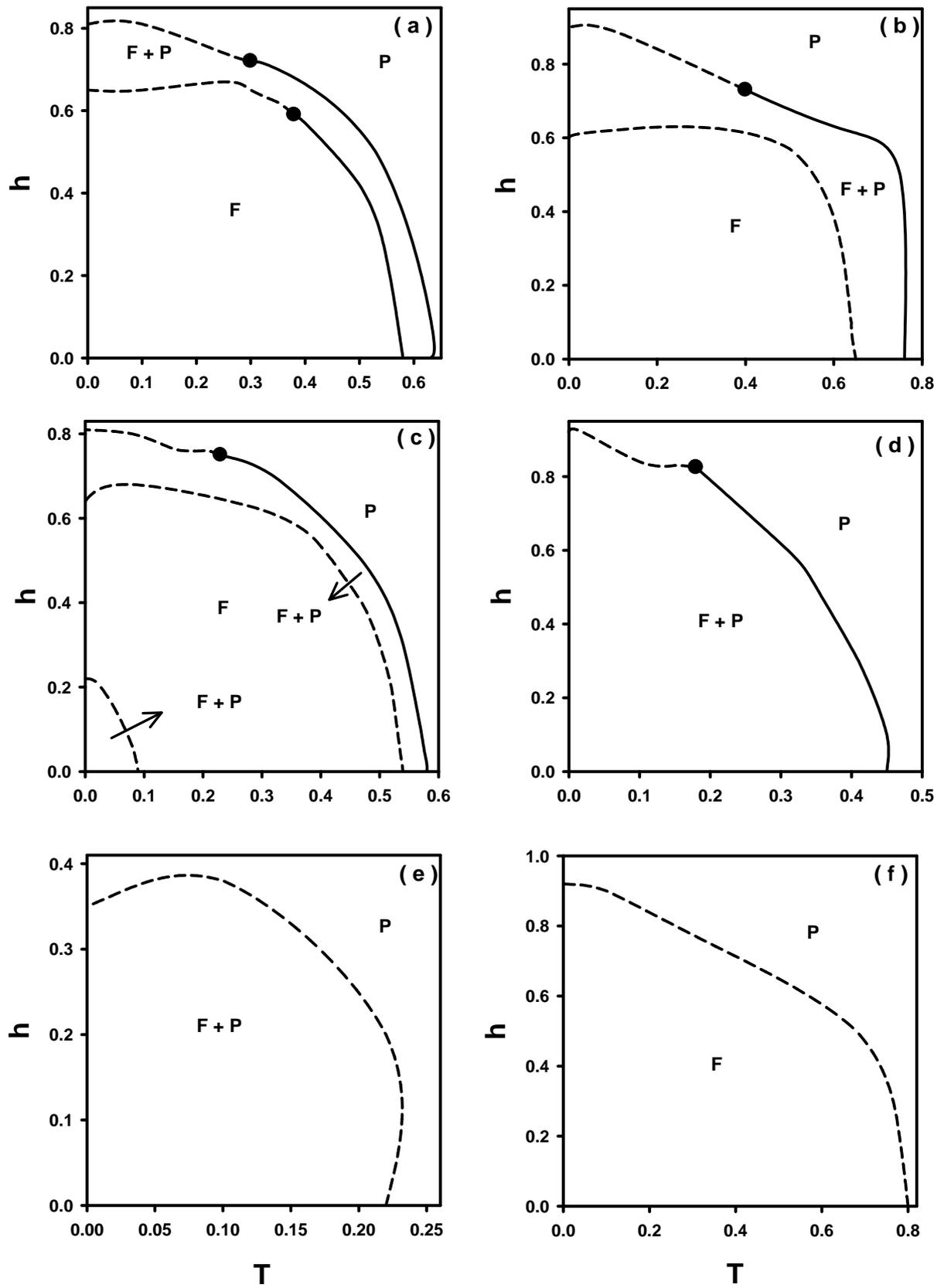

**FIG. 3**

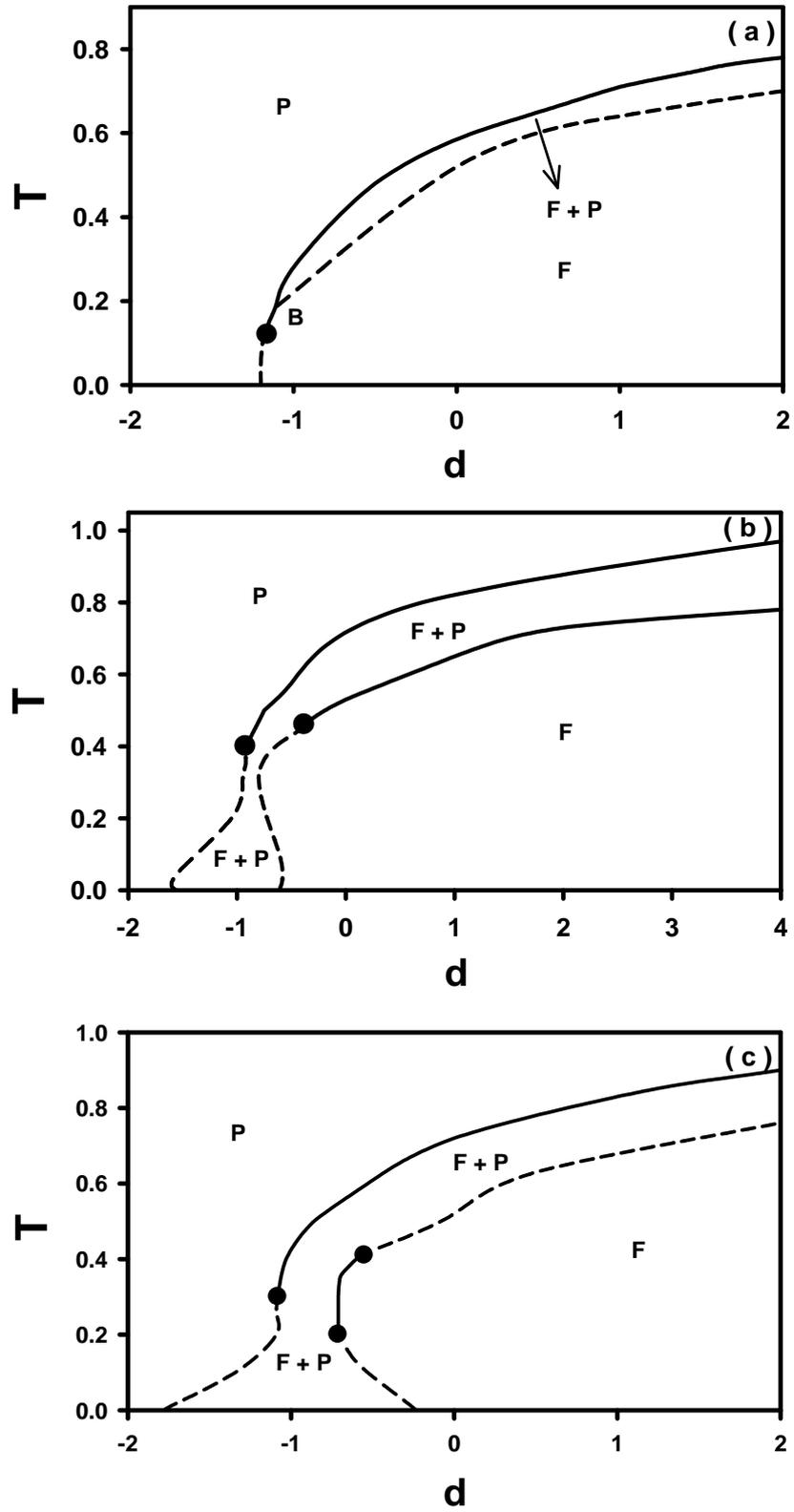

**FIG. 4**

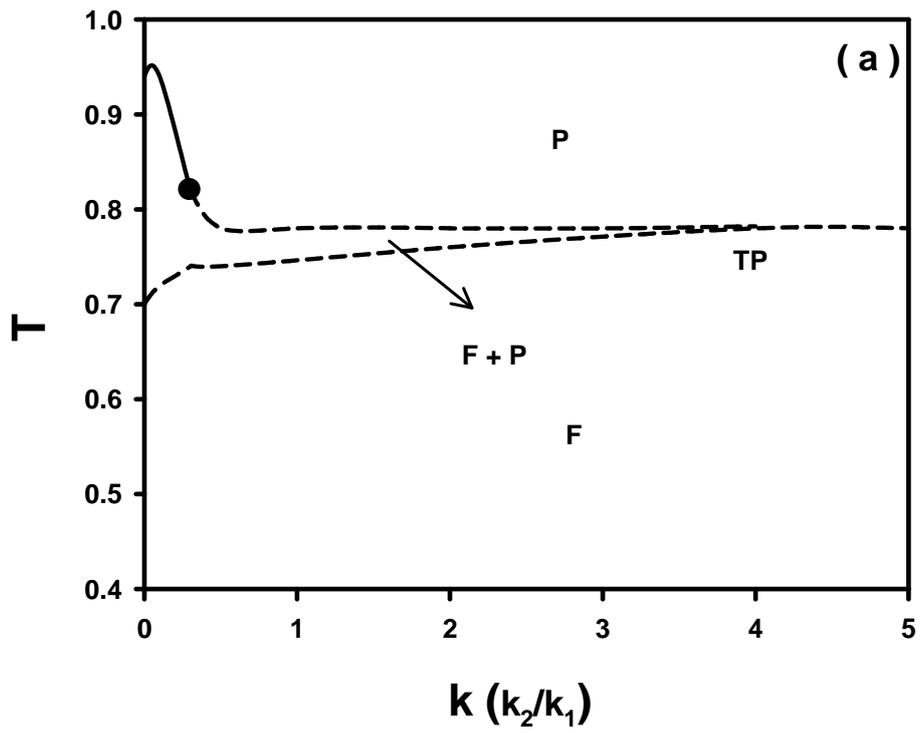

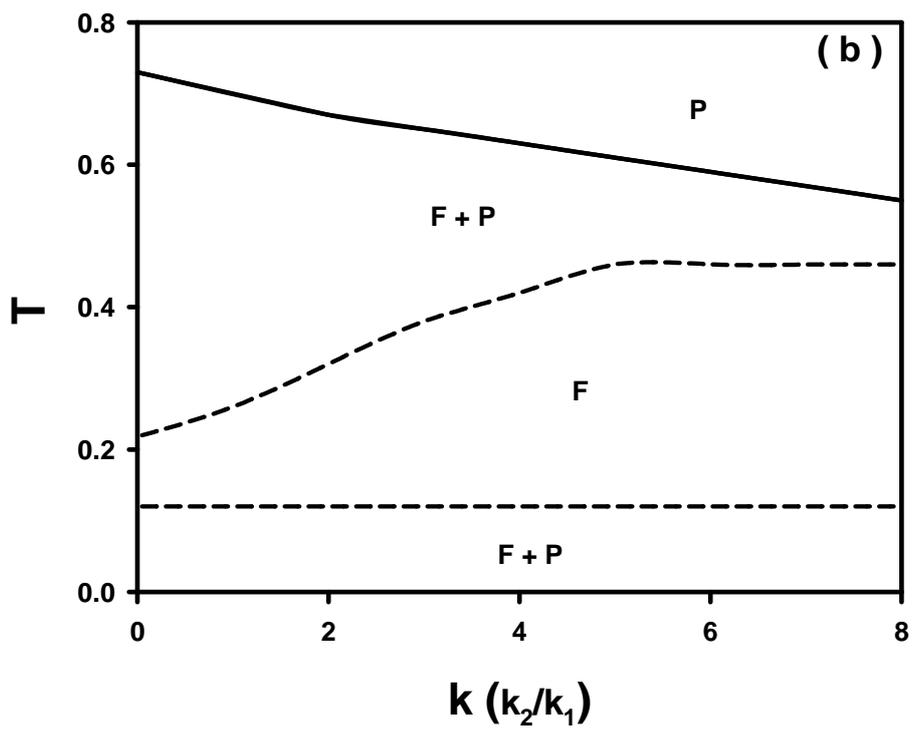

Fig. 5

|       | $X_1$ | $X_2$ | $X_3$ |
|-------|-------|-------|-------|
| $X_1$ |       | $k_1$ | $k_2$ |
| $X_2$ | $k_1$ |       | $k_1$ |
| $X_3$ | $k_2$ | $k_1$ |       |

**TABLE 1**

|        | $X_1$ | $X_2$ | $X_3$ |
|--------|-------|-------|-------|
| $X_1$  |       | $k_1$ | $k_2$ |
| $X_2$  | $k_1$ |       | $k_1$ |
| $X_3$  | $k_2$ | $k_1$ |       |

**TABLE 1**